\documentclass[prd,preprint,superscripointaddress,nofootinbib,longbibliography,aps]{revtex4-1}
\usepackage[utf8]{inputenc}
\usepackage{amsmath}
\usepackage{graphicx}
\usepackage{longtable}
\usepackage{calc}
\usepackage{accents}
\usepackage{color}
\usepackage{amssymb}
\usepackage{CJK}
\usepackage{subfigure}
\usepackage{microtype}
\usepackage[linktoc=all]{hyperref}
\usepackage{cleveref}
\usepackage{stackengine}
\usepackage{color}
\usepackage{bm}
\usepackage{braket}
\usepackage{slashed}
\usepackage{feynmf}
\usepackage{comment}
\usepackage{bbding}
\definecolor{myblue}{rgb}{ 0.188, 0.478,0.858}

 \newcommand{\dbtilde}[1]{\accentset{\approx}{#1}}

\begin{document}

\begin{CJK*}{UTF8}{gbsn}

\preprint{Imperial/TP/2022/MC/01}
\title{Double Copy for Massive Scalar Field Theories}

\author{Mariana Carrillo Gonz\'alez$^{(1)}$} \email{m.carrillo-gonzalez@imperial.ac.uk}
\author{Qiuyue Liang (梁秋月)$^{(2)}$} \email{qyliang@sas.upenn.edu}
\author{Mark Trodden$^{(2)}$} \email{trodden@physics.upenn.edu}
\affiliation{$^{1}$Theoretical Physics, Blackett Laboratory, Imperial College, London, SW7 2AZ, U.K \\ $^{2}$Center for Particle Cosmology, Department of Physics and Astronomy, University of Pennsylvania, Philadelphia, Pennsylvania 19104, USA}

\begin{abstract}
We explore extensions of the double copy to massive theories and find a new cubic theory with a local double copy. We consider the nonlinear sigma model and the special galileon theory, massless versions of which are known to be related through the double copy. We show that by performing a Kaluza-Klein reduction of these theories from five dimensions down to four, a double copy relation exists between the resulting massive four-dimensional scalar field theories. This requires the vanishing contribution of new galileon terms arising in high dimensions. We further explore if other interactions that do not arise from a dimensional reduction of the nonlinear sigma model could be double copied and find a new cubic interaction which satisfies the BCJ relations up to 5-point amplitudes.
\end{abstract}

\maketitle

%%%%%%%%%%%%%%%%%%%%%%%%%
\section{Introduction}
%%%%%%%%%%%%%%%%%%%%%%%%%
It is now quite firmly established that there exists a web of relationships among the scattering amplitudes of different theories, suggesting deep underlying connections relating seemingly very different models, describing what is thought to be very different physics. First seen in string theory as a map between closed and open string amplitudes~\cite{Kawai1986}, the relationship has been extensively explored, in the field theory limit, as one between Yang-Mills (YM) theories and gravity and is referred to as {\it the double copy}. For example, the Kawai-Lewellen-Tye (KLT) double copy \cite{Kawai1986} is given by 
\begin{equation}
\label{eq,doublecopy}
    \mathcal{A}_{n}^{A \otimes B}(1,2, \cdots, n)=\sum_{\alpha, \beta} \mathcal{A}_{n}^{\mathrm{A}}[\alpha] S[\alpha|\beta] \mathcal{A}_{n}^{\mathrm{B}}[\beta], 
\end{equation}
where $\mathcal{A}_n^{I}$ is the n-point amplitude partial amplitude for the theory $I$. $\alpha, \beta$ denote the color ordering and $S[\alpha|\beta]$ is the KLT kernel. Beyond the possibility that this relationship reveals deep connections among our fundamental theories, the double copy is also proving to be an important practical advance, since it holds the promise of simplifying gravitational computations that are extremely complex due to the nonlinearity of gravity. 

In recent years the double copy has been generalized in a range of different directions including to scalar effective field theories such as the non-linear sigma model (NLSM) and the Special Galileon (SGal), to theories with supersymmetry, to gauged supergravities, Yang-Mills Einstein theory, Born-Infeld theory, and to several other examples \cite{Bern:2019prr}. To understand what underlies these various implementations, it is crucial to study the key ingredients of the double copy relation: color-kinematics duality, and the Bern-Carrasco- Johansson (BCJ) relation \cite{Bern:2008qj}. Furthermore, various modifications of the double copy have aroused interest, and as a bottom-up effective field theory approach, it is interesting to write down the most general amplitudes that can satisfy color-kinematics duality and lead to a well-behaved double copy. Exploring different ways to generalize the double copy relation is thus important if we are to understand the origin of these relationships. There are several directions in which one might consider modifying the double copy. One interesting direction is to encode the higher derivative corrections into a modified momentum-dependent color factor which satisfies the same algebra as the original color factors \cite{Carrasco:2019yyn,Low:2019wuv,Low:2020ubn,Carrasco:2021ptp}. Another is to generalize the double copy starting from a modified form of the KLT kernel to encode the higher derivative terms \cite{Chi:2021mio,Mizera:2016jhj}. 

In this paper, we will focus on a particular generalization consisting of extending the double copy to massive theories~\cite{Johnson:2020pny,Momeni:2020hmc,Momeni:2020vvr,Gonzalez:2021bes,Gonzalez:2021ztm,Moynihan:2020ejh,Moynihan:2021rwh,Burger:2021wss,Hang:2021fmp,Hang:2021oso}. For a generic massive theory with a single state, the massive KLT kernel matrix is full-rank. Therefore, as has been explicitly checked up to six point amplitudes~\cite{Johnson:2020pny}, the matrix always has an inverse. On the other hand, for a massless theory, the KLT kernel has reduced rank, and thus is not invertible, leading to the existence of null vectors. Each null vector results in $(n - 3)(n - 3)!$ BCJ relations as extra constraints on the basis of partial amplitudes. The absence of  null vectors in the massive case indicates that there are no extra BCJ-like constraints in such theories. This leads to a double copy with spurious poles staring at 5-point amplitudes. One possible way of cancelling these  spurious poles is to require a tower of massive fields satisfying the {\it mass spectral condition}~\cite{Johnson:2020pny}, which reduces the rank of the KLT kernel to be that of the massless case. This enables us to write down the null vector of the massive KLT kernel and the analog of the massless BCJ relations. An example of a theory satisfying the mass spectral condition is the Kaluza-Klein reduction of Yang-Mills theory. In fact, if one considers a gauge theory with the same operators as those arising in the Kaluza-Klein theory, but with arbitrary coefficients, color-kinematics duality fixes these coefficients to be those of the Kaluza-Klein theory \cite{Momeni:2020hmc}. Other massive theories arising from a dimensional reduction have been studied in the context of the double copy in \cite{Chiodaroli:2015rdg,Chiodaroli:2017ehv,Chiodaroli:2018dbu,Bautista:2019evw}.

It is natural to ask whether all possible 4D theories that satisfy color-kinematics duality and lead to a well-defined double copy arise from the Kaluza-Klein compactification of a  higher dimensional massless theory. To investigate this question, we will explore the double copy relation among massive scalar theories. In order to have a local double copy we focus on the case of a tower of massive scalar fields that satisfy the mass spectral conditions.  We will explore the dimensional reduction of the 5D massless nonlinear sigma-model and special galileon theory, and will investigate whether there are other possible interactions satisfying color-kinematics duality.

The paper is organized as follows. In Sec II, we analyze the Kaluza-Klein compactification of higher dimensional massless theories. We discuss how the mass spectral condition is automatically satisfied and introduce a new set of Mandelstam variables that correspond to the dimensional reduction of the 5 dimensional (5D) variables. In Sec III, we discuss the most general 4-point amplitude that can satisfy the Kleiss-Kluijf (KK) \cite{Kleiss:1988ne} and BCJ relations\cite{Bern:2008qj}. We find that this is identical to the one found in the dimensionally-reduced massless theory. In Sec IV, we discuss the cubic interactions that can satisfy the KK and BCJ relations. We conclude that as long as the 3-point amplitude is cyclic and anti-symmetric, the KK relation is automatically satisfied both for the 4-point and 5-point amplitudes. For a constant 3-point amplitude, once the 4-point BCJ relation is satisfied, the fundamental 5-point BCJ relation is automatically satisfied, but other 5-point BCJ relations represent extra constraints. We show that these constraints can always be satisfied if we consider an infinite tower of massive states by counting the free parameters in the theory. We conclude in Sec V. 

\section{Massive scalar double copy from Kaluza-Klein reduction} 
\subsection{NLSM}
In this section, we will show that the Kaluza-Klein compactification of the 5d non-linear sigma model (NLSM) over an $S^1$ automatically satisfies the mass spectral condition in 4 dimensions (4D). We begin by discussing the massless NLSM in 5D. The NLSM describes the symmetry breaking pattern $G_L \times G_R \to G$, where $G_{L,R}$ are compact Lie groups. Here, we consider the case $G_L=G_R=G=U(N)$ in order to have well-defined color ordered amplitudes. To obtain the effective Lagrangian to all derivative orders, it is convenient to use the coset construction. The lowest derivative order Lagrangian reads
\begin{equation}
    \mathcal{L} = \frac{F^2}{4} \text{Tr}\left(\partial_\mu \mathcal{U}^{\dagger} \partial^\mu \mathcal{U} \right) \ ,
\end{equation}
where $F$ is the symmetry breaking scale and $\mathcal{U}$ is an element of the group $U(N)$. In the Cayley parametrization, one has 
\begin{equation}
    \label{eq,cayley}
    \mathcal{U} = 1+2 \sum_{n=1}^{\infty}\left(\frac{i}{2 F} \hat \phi\right)^{n}\ ,
\end{equation}
where $\hat\phi =\hat\phi^a t^a= \hat\phi(x^\mu, y)$, $\hat\mu = (\mu,y)$, $t^a$ are the generators of $U(N)$, and our notation is such that objects with a hat are 5D quantities. Greek letters $\mu , \nu , \ldots$ denote 4D indices and $y$ denotes the 5th dimension. The leading order term in $\phi$ corresponds to the kinetic term,
\begin{equation}
   \mathcal{L}_{\mathrm{NLSM}}^{(2)} = \frac{1}{2} \partial_{\hat\mu} \hat\phi  \partial^{\hat\mu} \hat\phi ~.
\end{equation}
Suppose the space has topology $R^4 \times S^1$, so that the coordinate $y$ is periodic, $ 0\leq m y < 2\pi$, where $m$ is the inverse radius of the circle $S^1$. We can then express the field as
\begin{equation}
    \hat \phi = \sum_{n = -\infty}^{n = \infty} \phi_n (x^\mu) e^{ i n m y} ,\quad \phi_n^{*}(x)=\phi_{-n}(x) \ ,
\end{equation}
\begin{comment}
\begin{equation}
 \partial_{\hat \mu} \hat \phi = \left\{
             \begin{array}{lr}
             \sum  \partial_\mu \phi_n (x^\mu) e^{ i n m y} , &  \hat \mu = \mu= (0,1,2,3)\\
             \sum ( i n m) \phi_n (x^\mu) e^{ i n m y}  , &  \hat \mu = y
             \end{array}
\right.
\end{equation}
\end{comment}
so that the kinetic term reads\thefootnote{We use metric signature $(+ - - - -)$}
\begin{eqnarray}
    \partial_{\hat \mu} \hat \phi\partial^{\hat \mu} \hat \phi 
    &=&  \partial_{\mu} \hat \phi\partial^{\mu} \hat \phi -  \partial_{y} \hat \phi\partial_{y}  \hat \phi =
    \sum_{n_1,n_2} e^{i (n_1+n_2) m y} \left( \partial_\mu \phi_{n_1}\partial^\mu \phi_{n_2} + n_1 n_2 m^2 \phi_{n_1}\phi_{n_2}\right) \ .
\end{eqnarray}
Integrating over $y$ from $0$ to $2\pi/m$ to obtain the 4D Lagrangian then gives:
\begin{equation}
    \mathcal{L}^{(2)}_{4D } = \frac{1}{2} \times \left(\frac{2\pi}{m}\right) \sum_{n_1, n_2} \delta(n_1 + n_2) \left(\partial_{ \mu}  \phi_{n_1}  \partial^{\mu}\phi_{n_2} + n_1 n_2 m^2 \phi_{n_1}\phi_{n_2} \right) \ .
\end{equation}
For simplicity, we will set $m = 2\pi$ here; this is equivalent to redefining the field to make the kinetic term canonical and setting it to have dimensions of mass as it should have in 4D. We further assume that the fields are all real, so that $\phi_{-n } = \phi_{n}$. Thus, we arrive at our first observation: $m^2_{\phi_n} = n^2 m^2$, which we will later denote as $m_n^2 = n^2 m^2$.

To study the 4-point amplitude, we look at the quartic interaction in the 5D NLSM,
\begin{eqnarray}
      \mathcal{L}^{(4)} &=& -\frac{1}{2F^2} \text{Tr} (t^at^b t^c t^d )\partial_{\hat\mu} \hat\phi^a  \partial^{\hat\mu} \hat\phi^b \hat \phi^c \hat \phi^d    \nonumber\\
    &=&  -\frac{1}{2F^2} \text{Tr} (t^at^b t^c t^d ) \sum_{n_1,2,3,4} e^{i (n_1 +\cdots n_4) m y} \left( \partial_\mu \phi^a_{n_1}\partial^\mu \phi^b_{n_2} + n_1 n_2 m^2 \phi^a_{n_1}\phi^b_{n_2} \right) \phi^c_{n_3}\phi^d_{n_4} ~.
\end{eqnarray}
After integrating over the fifth dimension, we have
\begin{eqnarray}
      \mathcal{L}^{(4)}_{4D} &=& 
   -\frac{1}{2F^2} \frac{2\pi}{m} \text{Tr} (t^at^b t^c t^d ) \sum_{n_1,2,3,4}\delta(n_1+\cdots n_4)  \left( \partial_\mu \phi^a_{n_1}\partial^\mu \phi^b_{n_2} + n_1 n_2 m^2 \phi^a_{n_1}\phi^b_{n_2} \right) \phi^c_{n_3}\phi^d_{n_4} \ .
\end{eqnarray}
Notice that the momentum space Feynman rule for the term $\left( \partial_\mu \phi_{n_1}\partial^\mu \phi_{n_2} + n_1 n_2 m^2 \phi_{n_1}\phi_{n_2} \right)$ $ \phi_{n_3}\phi_{n_4} $ is $ - i \left(p_1 \cdot p_2 - n_1 n_2 m^2 \right)$. We can use the on-shell condition for the external legs, $p_i^2 = m_i^2 = n_i^2 m^2$, to simplify the vertex and obtain $- \frac{i}{2} \left((p_1 + p_2)^2 - (n_1 + n_2 )^2m^2\right)$. It is therefore useful to define the {\it mediating masses},
\begin{equation}
    m_{12}^2 \equiv (n_1 + n_2)^2 m^2, \quad m_{13}^2 \equiv (n_1 + n_3)^2 m^2,\quad m_{14}^2 \equiv (n_1 + n_4)^2 m^2~ , \label{eq,mediatingM}
\end{equation}
and a set of {\it tilded Mandelstam variables}
\begin{equation}
\label{eq,newMandel}
    \tilde s_{ij} \equiv s_{ij}- m_{ij}^2 ~,
\end{equation}
with $s_{ij}$ the standard Mandelstam variables
\begin{equation}
    s = s_{12}=(  p_1+  p_2)^2,\quad t =s_{13}= (  p_1+  p_3)^2,\quad u =s_{14}= (  p_1+  p_4)^2 ~,
\end{equation}
where all the momenta are outgoing. The Feynman rule for a single term can then be simplified as $-\frac{i}{2} \tilde s_{12}$. To obtain the 4-point color-ordered amplitude for such a term, we must sum over all cyclic permutations:
\begin{equation}
       \mathcal{A}_4[1234]=  \frac{1}{4F^2} \left(\tilde s_{12} + \tilde s_{23} + \tilde s_{34} + \tilde s_{41} \right) = - \frac{1}{2F^2} \tilde s_{13} \ ,
\end{equation}
where we have used the following identity
\begin{equation}
\label{eq,moment4point}
    \tilde s_{12}+ \tilde s_{13}+ \tilde s_{23} =\sum_{i=1}^4 m_i^2 - m_{12}^2 -m_{13}^2 -m_{14}^2 =0~,
\end{equation}
which is obtained from momentum conservation and the constraint arising from the Kaluza-Klein reduction, $\delta(n_1 + n_2 +n_3 +n_4)$. The last equality is the mass spectral condition pointed out in \cite{Johnson:2020pny}, which is automatically satisfied for the massive tower of states arising from the Kaluza-Klein reduction. The 4-point color ordered amplitude in the massive theory is, therefore, equal to that of the massless theory after the replacement $s_{13} \rightarrow \tilde s_{13}$. This property is satisfied for all n-point scattering amplitudes of a Kaluza-Klein theory, that is,
\begin{equation}
    \mathcal{A}^{\text{K.K}}(\tilde s_{ij})= \mathcal{A}^{\text{massless}}(s_{ij}\rightarrow \tilde s_{ij}) \ , \label{eq,kkamplitude}
\end{equation}
where $s_{ij}$ are the 5D Mandelstam variables and $\tilde s_{ij}$ are the dimensionally reduced versions. This simply follows from the dimensional reduction of the 5D momenta: $\hat{p}_i=(p_i,m_i)$, the on-shell condition of the external legs, and the definition of the Kaluza-Klein and mediating masses. 

It is worthwhile computing the 6-point amplitude in the NLSM to see explicitly how the amplitude is indeed a function of $\tilde s_{ij}$. The 6-pion Lagrangian reads
\begin{equation}
    \mathcal{L}^{(6)} = 
    \frac{1}{8F^4} \operatorname{Tr}\left(t^{a} t^{b} t^{c} t^{d}t^{e}t^{f}\right)\left( 2 \partial_{\hat{\mu}} \hat{\phi}^{a} \partial^{\hat{\mu}} \hat{\phi}^{b} \hat{\phi}^{c} \hat{\phi}^{d}  \hat{\phi}^{e} \hat{\phi}^{f} + \partial_{\hat{\mu}} \hat{\phi}^{a} \hat{\phi}^{b} \partial^{\hat{\mu}}  \hat{\phi}^{c} \hat{\phi}^{d}  \hat{\phi}^{e} \hat{\phi}^{f} \right) \ , \label{eq,6pointNLSM}
\end{equation}
so that the amplitude is
\begin{equation}
    \mathcal{A}_6 [123456] = \frac{1}{2F^4} \left( \frac{\tilde s_{13 }\tilde s_{46}}{\tilde s_{123}}+\frac{\tilde s_{15 }\tilde s_{24}}{\tilde s_{156}}+\frac{\tilde s_{26 }\tilde s_{35}}{\tilde s_{126}} - \tilde s_{24} -\tilde s_{46} - \tilde s_{26} \right) \ ,
\end{equation}
where the first three terms come from a pole term with two quartic vertices, and the last three terms come from the sextic contact term. This is the same as the 6-point amplitude in the massless case, with the tilded Mandelstam variables replacing the standard ones. 

It has been shown that the mass spectral condition guarantees the cancellation of spurious poles in the KLT kernel at 5-points, therefore maintaining locality in the double copy. In the 5D NLSM there are no 5-point amplitudes, since we cannot write a Wess-Zumino term; nevertheless, we expect this condition to be relevant at higher orders as well. Note that at 4-points the introduction of the mediating masses is just an artifact that allows us to rewrite the contact term in a simplified form, but at 6-points this indeed corresponds to mediating masses which for every 4-point sub-graph satisfy the spectral condition. We define $\tilde s_{ijk } \equiv (p_i+ p_j+p_k)^2 - m_{ijk}^2$, where in a 4-point scattering process, $\tilde s_{123} = p_4^2 - m_4^2 =0 $ since the 4th leg is on-shell, in a 5-point scattering process, $\tilde s_{123} = \tilde s_{45} $, and in a 6-point scattering process, $\tilde s_{123} = (p_{1}+p_2+p_3)^2 - m_{123}^2 $ is the internal line inverse propagator. The tilded Mandelstam variables then satisfy 
\begin{equation}
    \tilde s_{ij} +  \tilde s_{jk} + \tilde s_{ki} =  \tilde s_{ijk}   ~,
\end{equation}
as in the massless case. In fact, the mass spectral condition ensures that the tilded Mandelstam variables satisfy the same relations as the standard Mandelstam variables. Therefore the massive KLT kernel has the same rank as in the massless case and does not contain spurious poles for all n-point amplitudes.

\subsection{SGal}
The Special Galileon (SGal) is a low energy effective field theory that non-linearly realizes a spacetime-dependent shift symmetry. One can build up the most general Lagrangian using the methods introduced in \cite{Hinterbichler:2015pqa,CarrilloGonzalez:2019fzc,Garcia-Saenz:2019yok}. To obtain the tower of massive special galileon states, we apply the same Kaluza-Klein reduction procedure as for the NLSM.  We start with the following Lagrangian 
 \begin{eqnarray}
 \label{eq,SGalL}
      \mathcal{L}_{\text {SGal }}&=&\frac{1}{2}(\hat\partial \hat\pi)^{2}-\frac{\alpha }{12}(\hat\partial \hat\pi)^{2}\left[(\hat\square \hat\pi)^{2}-\left(\partial_{\hat\mu} \partial_{\hat\nu} \hat\pi\right)^{2}\right] \nonumber\\
     &+& \frac{\alpha^2 }{240}(\hat\partial \hat\pi)^{2}  \left[(\hat\square \hat\pi)^4 - 6(\hat\square \hat\pi)^2 \left(\partial_{\hat\mu} \partial_{\hat\nu} \hat\pi\right)^{2} + 8 \hat\square \hat\pi \left(\partial^{\hat\mu} \partial_{\hat\nu} \hat\pi \partial^{\hat\nu} \partial_{\hat\rho} \hat\pi \partial^{\hat\rho} \partial_{\hat\mu} \hat\pi\right)\right. \nonumber\\
     &&~~~~ \left. + 3 \left(\partial_{\hat\mu} \partial_{\hat\nu} \hat\pi\right)^{2} \left(\partial_{\hat\rho} \partial_{\hat\sigma} \hat\pi\right)^{2} - 6 \left(\partial^{\hat\mu} \partial_{\hat\nu} \hat\pi \partial^{\hat\nu} \partial_{\hat\rho} \hat\pi \partial^{\hat\rho} \partial_{\hat\sigma} \hat\pi\partial^{\hat\sigma} \partial_{\hat\mu} \hat\pi\right)   \right] \ ,
 \end{eqnarray}
 where $\hat \partial = \partial_{\hat \mu}$ is the 5D derivative. The terms with six Galileons first appear in 5D (see \cite{Hinterbichler:2015pqa}), and are absent in lower dimensions. The Kaluza-Klein reduction of the kinetic term follows in a similar manner to the NLSM case, so that the masses are again given by $m_{\pi_{n_i}}^2=m_i^2 =n_i^2 m^2 $. 
 
The quartic term contains two pieces. After Kaluza-Klein reduction the $(\hat\square \hat \pi)^2$ term vanishes as long as one leg is on-shell, and the $\left(\partial_{\hat{\mu}} \partial_{\hat{\nu}} \hat{\pi}\right)^{2}$ term reduces to
\begin{equation}
   ( \partial_\mu \partial_\nu \pi)^2 + 2 n_3 n_4 m^2 (\partial_\mu \pi) ^2 + n_3^2 n_4^2 m^2  \pi^2 \ ,
\end{equation}
in 4D. Thus, the 4-point amplitude is given by
\begin{eqnarray}
    \mathcal{A}_{\text{SGal}4}&= &  \left( -\frac{\alpha}{12}  \right)   \left(\tilde{s}^3+\tilde{t}^3+\tilde{u}^3 \right)=  -\frac{\alpha}{12 }  \left(3 \tilde{s}\tilde{t}\tilde{u}  \right)  \ .
\end{eqnarray}
One can immediately see that the double copy relation (Eq.\eqref{eq,doublecopy}) between the NLSM and the SGal 4-point amplitudes is satisfied for the Kaluza-Klein setup in the same manner as in the massless case. In fact, given the observation in Eq. \eqref{eq,kkamplitude}, we reach the conclusion that the double copy relation between the n-point amplitudes of the leading derivative order NLSM and SGal Kaluza-Klein theories compactified from higher dimensional massless theories follows directly from the massless case. The double copy is  trivially realized since the massive tower of states always satisfies the spectral condition.

We now proceed to study the new 6-point amplitude arising from the sextic Galileon contact term. As discussed before, the contribution from the $\hat\square \hat\pi$ term will vanish after the Kaluza-Klein reduction when using the on-shell condition. Therefore, only two terms contribute to the 6-point amplitude: the first one is similar to the quartic interaction case, and contributes to the amplitude with a term proportional to
\begin{equation}
  3(\hat\partial \hat\pi)^{2}\left(\partial_{\hat{\mu}} \partial_{\hat{\nu}} \hat{\pi}\right)^{2}\left(\partial_{\hat{\rho}} \partial_{\hat{\sigma}} \hat{\pi}\right)^{2}  \to  \frac{3}{16} \tilde s_{12} \tilde s_{34}^2 \tilde s_{56}^2\ ,
\end{equation}
while the other term gives,
{\small
\begin{align}
   -&6(k_1\cdot k_2 - n_1 n_2 m^2 ) (k_3\cdot k_4 - n_3 n_4 m^2 )(k_4\cdot k_5- n_4 n_5 m^2)( k_5\cdot k_6- n_5 n_6 m^2 )( k_6\cdot k_3    - n_6 n_3 m^2) \nonumber \\
   &= -\frac{3}{8} \tilde s_{12} \tilde s_{34} \tilde s_{45} \tilde s_{56} \tilde s_{63 } \ . 
\end{align}
}
The 6-point amplitude from this new contact term is $\frac{\alpha^2}{240} \frac{\tilde s_{12}}{4} \left(\frac{3}{16} \tilde s_{34}^2 \tilde s_{56}^2 - \frac{3}{8} \tilde s_{34} \tilde s_{45}\tilde s_{56}\tilde s_{63} \right)$ $ +\text{ permutations}$. We have checked numerically that summing over all permutations yields a vanishing amplitude. This reaffirms Eq.\eqref{eq,kkamplitude}, which tells us that the 4D massive amplitude is simply the massless amplitude with the substitution $s_{ij} \rightarrow \tilde s_{ij}$, since the extra term from 5D vanishes in 4D.  

We would now like to address the question of whether there are other 4D massive theories that can satisfy the double copy relation, but which do not come from a higher-dimensional reduction. In \cite{Momeni:2020hmc}, the authors considered operators arising from the compactification of Yang-Mills over $S^1$, but left the coefficients arbitrary. By imposing the KK and BCJ relations on the color-ordered amplitudes, they concluded that the coefficients are fixed to be those arising from the Kaluza-Klein compactification. However, other types of operators might exist in a generic theory. In the following sections, we will analyze this possibility for the case of a massive tower of colored scalars that satisfy the spectral condition.

\section{Exploring quartic interactions}
\label{sec,quartic}
In this section we want to explore possible generalizations of 4-point amplitudes from contact terms by including the masses of the multi-scalar theory with a $U(N)$ symmetry as new building blocks. We focus on local theories so that the functions of Mandelstam variables are simple polynomials, and we assume that the mass spectral condition is satisfied. We will use the 4-point KK relation
\begin{equation}
\label{eq,4pointKK}
    \mathcal{A}_{4}[1234]+\mathcal{A}_{4}[2134]+\mathcal{A}_{4}[2314]=0 \ ,
\end{equation}
and the BCJ relation
\begin{equation}
\label{eq,4pointBCJ}
   \tilde u  \mathcal{A}_{4}[1234]=\tilde t  \mathcal{A}_{4}[1243] \ ,
\end{equation}
to constrain the color-ordered 4-point amplitude. 
 
We start by analyzing the lowest order, $\mathcal{O}(p^0)$, where we can only have constant coefficients. For generality, we allow these coefficients to carry an extra structure which indicates that the interaction strength can be different for different massive states, thus
\begin{equation}
\label{eq,4c0}
   \mathcal{A}_{4}[1234] = c^{(0)}[1234] \ .
\end{equation}
It is then possible to make a specific choice of these constants to ensure that the KK relation, Eq.\eqref{eq,4pointKK}, is satisfied. However, it is {\it not} possible for such a choice of coefficients to also satisfy the BCJ relation, Eq.\eqref{eq,4pointBCJ}, since $\tilde u$ and $\tilde t$ are independent variables. At $\mathcal{O}(p^2)$, things are different, since the amplitudes are now momentum dependent. We write the most general 4-point amplitudes as follows,
\begin{equation}
\label{eq,A4point1}
   \mathcal{A}_{4}[1234] = \frac{1}{F^2}\left(c_1[1234] ~\tilde s + c_2[1234] ~\tilde t \right) \ ,
\end{equation}
where $\tilde s $ and $\tilde t $ are the tilded Mandelstam variables as defined in Eq.\eqref{eq,newMandel}, and $c_1$ and $c_2$ are dimensionless constants. Inserting this expression into the KK relation, we obtain
\begin{eqnarray}
  \label{eq,4pointKKexpample1}
    && \left(c_1[1234]+ c_1[2134]+ c_2[2314] -c_2[1234] \right) \tilde s +  \left(c_1[2314]-c_2[1234] +c_2[2134] \right)  \tilde u =0 \ ,
\end{eqnarray}
where we have used momentum conservation. Since $\tilde s$ and $\tilde u$ are independent variables, the coefficients in front of them should vanish independently, and thus we have two constraints from the KK relation,
\begin{equation}
    c_1[1234]+ c_1[2134]+ c_2[2314] -c_2[1234] =0\  ,\quad c_1[2314]-c_2[1234] +c_2[2134]  =0 \ .
\end{equation}
The BCJ relation, on the other hand, gives us
\begin{equation}
    - \tilde u^2 c_1[1234] +\tilde t^2 c_1[1243] + \tilde u \tilde t \left(-c_1[1234]+ c_1[1243]+ c_2[1234]- c_2[1243] \right)  =0\ ,  \nonumber\\
\end{equation}
which leads us to,
\begin{eqnarray}
\label{eq,4pointBCJc123}
  c_1[1234] = c_1[1243] = c_2[1234]- c_2[1243] =0 \ ,
\end{eqnarray}
that is, that $c_1$ should vanish and the $c_2$ coefficients should be symmetric among particles $3$ and $4$. Using these constraints together with the KK relation Eq.\eqref{eq,4pointKKexpample1}, we find
\begin{equation}
    c_2[2314]- c_2[1234] =0, \quad c_2[2134]-c_2[1234] =0~.
\end{equation}
The latter expression tells us that $c_2$ is also symmetric among particle $1$ and particle $2$. Using this symmetry we can rewrite the former relation as
\begin{equation}
     c_2[2314]- c_2[1234] = c_2[2314]- c_2[2134] =0~.
\end{equation}
Therefore, we conclude that $c_2[1234] = c$ is just a constant without any particle labelling; i.e. that the strength of the interaction is the same for all massive states. Thus, the allowed 4-point amplitude starting from the most general construction in Eq.\eqref{eq,A4point1} reduces to the one obtained from a Kaluza-Klein reduction of the NLSM.

One could continue on to higher-derivative orders, $\mathcal{O}(p^n)$, explicitly checking whether all the allowed terms are equivalent to those arising from the Kaluza-Klein compactification of the NLSM. However, a general argument now seems clear. Once the mass spectral condition is satisfied, the tilded Mandelstam variables defined in Eq.\eqref{eq,newMandel} enjoy the same relations as the original Mandelstam variables. Moreover, since a constant amplitude cannot satisfy the BCJ relations, the amplitudes in the massive theory that satisfy both the KK and BCJ relations are those of the massless theory in which the Mandelstam variables are exchanged for their tilded companions.

In the remainder of the paper, we will therefore focus on a different question: whether a multi-scalar theory, in which the spectral relation is satisfied, can include a cubic interaction and serve as an example of a theory with a sensible double copy that does {{\it }not} arise from the Kaluza-Klein reduction of the NLSM.  

\section{A new cubic interaction}
We start by writing down the most general color-ordered cubic interaction $V_3[n_1,n_2,n_{12}]$, where we use the mass, $m_{n_1}=n_1 m$, to label the particles. The Lagrangian for such a term reads,
\begin{equation}
    \mathcal{L} \supset \text{Tr} [t^a t^b t^c] V_3[n_1,n_2,n_{12}] \phi^a _{n_1}\phi^b _{n_2}\phi^c _{n_3} \ . \label{eq:3point}
\end{equation}
By requiring cyclicity $V_3[n_1,n_2,n_{12}] = V_3[n_2,n_{12},n_1]= V_3[n_{12},n_1,n_2]$ and reflection $V_3[n_1,n_2,n_{12}] = (-1)^3 V_3[n_{12},n_2,n_1]$, one can see the cubic amplitude must be anti-symmetric. We then construct the 4-point on-shell color-ordered amplitudes with a color dressing $\rm{Tr}[t^{a_1} t^{a_2} t^{a_3} t^{a_4}]$ based on this vertex. We define the {\it one-leg off-shell} cubic amplitude as
\begin{equation}
    \tilde V_3[n_1,n_2,n_{12}] = V_3[n_1,n_2,n_{12}]  +  \frac{a_1[n_1,n_2,n_{12}] }{F^2}\tilde s_{12}+  \frac{a_2[n_1,n_2,n_{12}] }{F^4}\tilde s_{12}^2   + \cdots   \ ,
\end{equation}
where $\tilde s_{12} \to 0$ when taking all legs on-shell, so that $V_3[n_1,n_2,n_{12}]$ is the constant on-shell cubic amplitude. Notice that the cubic amplitude has mass dimension 1, so $\tilde V_3$, $V_3$, and $a_i$ also have mass dimension 1. The color-ordered 4-point amplitude is, 
\begin{eqnarray}
  \mathcal{A}_4[1234] =- \frac{\tilde V_3[n_1,n_2,n_{12}]\tilde V_3[n_3,n_4,n_{34}]}{\tilde s_{12}}-  \frac{\tilde V_3[n_4,n_1,n_{14}] \tilde V_3[n_2,n_3,n_{23}]}{\tilde s_{14}} \ ,
\end{eqnarray}
where the first term comes from the $s$ channel, and the second term from the $u$ channel. One can analyze this amplitude order by order in momentum, starting with the lowest order $\mathcal{O} (p^{0})$, where only the on-shell cubic vertex contributes. The 4 point KK relation Eq.\eqref{eq,4pointKK} leads to
\begin{eqnarray}
  && \frac{V_3[n_1,n_2,n_{12}] V_3[n_3,n_4,n_{34}]}{\tilde s_{12}}+ \frac{V_3[n_4,n_1,n_{14}] V_3[n_2,n_3,n_{23}]}{\tilde s_{14}}
  \nonumber\\
  &&+ \frac{V_3[n_2,n_1,n_{12}] V_3[n_3,n_4,n_{34}]}{\tilde s_{12}} + \frac{V_3[n_4,n_2,n_{24}] V_3[n_1,n_3,n_{13}]}{\tilde s_{13}} \nonumber\\
  &&  +\frac{V_3[n_2,n_3,n_{23}] V_3[n_1,n_4,n_{14}]}{\tilde s_{14}}+ \frac{V_3[n_4,n_2,n_{24}] V_3[n_3,n_1,n_{13}]}{\tilde s_{13}} =0  \ .
\end{eqnarray}
Requiring that the coefficient of $1/\tilde s_{12}$ vanishes requires $V_3[1,2,12] =- V_3[2,1,12]$, which is automatically satisfied, by construction, from reflection. Therefore, once cyclicity and reflection of the cubic amplitude is satisfied, the 4-point KK relation follows. The 4-point BCJ relation (Eq.\eqref{eq,4pointBCJ}) requires,
\begin{equation}
    - V_3[n_1,n_2,n_{12}] V_3[n_3,n_4,n_{12}] + V_3[n_1,n_3,n_{13}] V_3[n_2,n_4,n_{13}] - V_3[n_1,n_4,n_{14}]V_3[n_2,n_3,n_{14}] = 0 ~. \label{eq,4ptbcjcubic}
\end{equation}
This is a new example of an interaction that satisfies the BCJ relation and does not arise as the Kaluza-Klein reduction of the NLSM. It is worth noticing that after summing over all possible intermediate particle states, the 4-point BCJ relation we obtain above becomes the Jacobi identity of the structure constants of a Lie algebra. This indicates that the 3-point amplitude should be proportional to the structure constant of an infinite dimensional Lie algebra \footnote{One may consider the dimensional reduction of a biadjoint scalar (whose single color ordered amplitudes satisfy the KK and BCJ relations \cite{Bern:1999bx, Du:2011js}), but this cannot correspond to our example in Eq.~\eqref{eq:3point} since the global symmetries are finite-dimensional and this does not change under the Kaluza-Klein dimensional reduction.}. One option is to consider the algebra of (local) area-preserving diffeomorphisms on the sphere, where one can construct a map from the mass state $n_i$ to a 2-vector on the sphere  $\bm m (n_i) = ( m_1 (n_i)\ ,m_2 (n_i) ) $. The local algebra therefore takes the form 
\begin{eqnarray}
  [L_{\bm m}, L_{\bm l} ] = \epsilon^{ab} m_a l_b L_{\bm m+\bm l}\ , 
\end{eqnarray}
where $L_{\bm m} = -\mathrm{i} \epsilon^{a b} m_{a} \exp (\mathrm{i} m \cdot \boldsymbol{\sigma}) \partial_{b} $ is the local algebra generator, and $\bm \sigma = (\sigma_1, \sigma_2)$ are the coordinates on the sphere. This algebra is equivalent to the $\text{SU}_{+}(\infty)$ algebra \cite{Pope:1989cr}. The structure constants can serve as our 3-point amplitude:
\begin{equation}
V_3[n_1,n_2,n_{12}]=\epsilon^{ab} m_a(n_1) l_b(n_2) \delta_{n_1+n_2}^{n_{12}} \ .
\end{equation}
Similar algebras corresponding to volume-preserving diffeomorphisms have previously appeared as the kinematic algebras within double copy constructions \cite{Bjerrum-Bohr:2012kaa,Cheung:2022mix,Ben-Shahar:2021zww}. In our case, instead of encoding the momentum dependence of the amplitude, they encode the coupling strength between different massive states. It's worth noting that this is just one example of an additional internal symmetry that the infinite massive tower of scalars can have, but generically any infinite dimensional algebra whose structure constants are fully anti-symmetric will imply the BCJ relations.

In the following, instead of summing over all possible intermediate states we focus only on a finite subset. For this case, it is interesting to explore the free parameters and constraints in the theory. Suppose we choose a set of external particles to have masses $n_1, n_2, n_3, n_4$ and label them by $1,2,3,4$. For the 4-point amplitude with two cubic vertices, we have three more massive states with masses $n_{12}, n_{13}, n_{14}$ corresponding to intermediate particles and we label them by $5,6,7$. For the scattering process involving the external particles $\{1,2,3,4\}$, we have 7 mass states and 6 cubic amplitudes involved. We also have one mass spectral condition and one 4-point BCJ relation. Therefore, we are left with 11 free parameters: 6 masses and 5 cubic amplitudes. We further consider the case where the previous intermediate particle 
becomes an external state, i.e. the scattering processes between $\{i,j,k,l\}$, where $i,j=1,2,3,4$ and $k,l=5,6,7$. Assuming that only the initial 6 cubic amplitudes exist, we have 6 different such processes that lead to 6 new 4-point BCJ relations and 6 new spectral conditions. Imposing the new spectral conditions leads to the requirement of all states being massless.  We can add one more cubic amplitude between the states $5,6,7$ and the result will stay unchanged. This tells us that we cannot work with a finite number of massive states. Instead, we assume that there is a new interaction with a massive state, $p$, such that it can appear as a mediating mass in the $\{i,j,k,l\}$ scattering. For each of the 6 processes, we have 2 new vertices and 1 new mass, but also 1 new spectral condition and 1 new 4-point BCJ relation. This means that overall we have one new free parameter per process. Once again, we could try to look at the processes where $p$ is now an external state. Every time we do this, we need to add a new massive state so that we can satisfy the massive 4-point BCJ relations and spectral conditions. Since we have an infinite tower of massive states for each $\phi^a$, we will always be able to satisfy both the spectral conditions and the BCJ relations. This tells us that the mass spectral condition selects the scattering processes in the theory that should be grouped together to satisfy the 4-point BCJ relation. 

We now go to the next-to-leading order, where the off-shell leg contributes to the 4-point amplitude as
\begin{eqnarray}
  \label{eq,4point1}
    \mathcal{A}_4^{(1)} [1234] &= & - \frac{a_1[n_1,n_2,n_{12}]V_3[n_3,n_4,n_{34}]+V_3[n_1,n_2,n_{12}]a_1[n_3,n_4,n_{34}]  }{F^2} \nonumber\\
    && -  \frac{a_1[n_4,n_1,n_{41}]V_3[n_2,n_3,n_{23}]  +V_3[n_4,n_1,n_{41}]a_1[n_2,n_3,n_{23} ] }{F^2}
\end{eqnarray}
Once we require that the coefficients $a_1[n_1,n_2,n_{12}]$ satisfy cyclicity and reflection, the 4-point KK relation is satisfied for the amplitude above. Therefore, we could simplify the notation by denoting $a_1 [n_1,n_2,n_{12}] = V_3[n_1,n_2,n_{12}] a_1$. Note that the amplitude is constant, and thus if the cubic interaction is the only possible interaction in the theory, then satisfying the BCJ relation requires $a_1[n_1,n_2,n_{12}] =0 $. However, if we also allow for a 4-point interaction, we can use the contact term to cancel this. For example, we could choose the 4-point contact amplitude in Eq.\eqref{eq,4c0} to be the negative of Eq.\eqref{eq,4point1}. This leads to a vanishing 4-point amplitude, but it is necessary to check if this term can contribute to higher-point amplitudes. At next-to-next leading order, the two-cubic vertices 4-point diagram scales as $\mathcal{O}(p^2)$,
\begin{eqnarray}
  \label{eq,4point2}
   &&  \mathcal{A}_4^{(2)} [1234]=  -   \frac{ \tilde s_{12} }{F^4} \big(
a_1^2  V_3[n_1,n_2,n_{12}]V_3[n_3,n_4,n_{34}] +   a_2[n_1,n_2,n_{12}]V_3[n_3,n_4,n_{34}]  \nonumber\\
&+&  V_3[n_1,n_2,n_{12}]a_2 [n_3,n_4,n_{34}] \big)+ \frac{ \tilde s_{14} }{F^4} \big(
a_1^2  V_3[n_4,n_1,n_{14}]V_3[n_2,n_3,n_{23}]  \nonumber\\
&+&   a_2[n_4,n_1,n_{14}]V_3[n_2,n_3,n_{23}] + V_3[n_4,n_1,n_{14}]a_2 [n_2,n_3,n_{23}] \big) \ .
\end{eqnarray}
We find that $a_2[1,2,12]$ also has to satisfy cyclicity and reflection in order to make the 4-point KK relation work. The resulting 4-point amplitude can be simplified as 
\begin{equation}
     \mathcal{A}_4^{(2)} [1234]= -   \frac{   a_1^2 + 2 a_2  }{F^4} \left( \tilde s_{12} V_3[n_1,n_2,n_{13}]V_3[n_3,n_4,n_{34}] +\tilde s_{14} V_3[n_4,n_1,n_{14}]V_3[n_2,n_3,n_{23}]   \right)  \ ,
\end{equation}
and from the section above we already know that the only choice for the 4-point amplitude to satisfy the BCJ relation is to be proportional to $\tilde t $. Therefore, we either require $a_2 = - \frac{1}{2} a_1^2 $ or $V_3[n_1,n_2,n_{12}] V_3[n_3,n_4,n_{34}] = V_3[n_4,n_1,n_{14}]V_3[n_2,n_3,n_{23}]$. The latter possibility combined with the 4-point BCJ relation leads to $V_3[n_1,n_3,n_{13}]V_3[n_2,n_4,n_{13}] = 0$. Meanwhile, the former possibility sets the 4-point amplitude to zero but, as we will see next, it yields a non-zero 5-point amplitude.

We now move on to analyze the 5-point amplitudes. Since one of the three cubic vertices involved in this amplitude has two internal legs, we distinguish its cubic amplitude by defining the {\it two-leg off-shell} amplitude in the same spirit as the one-leg off-shell amplitude,
\begin{equation}
    \dbtilde V_3[n_1,n_{23},n_{45}] = V_3[n_1,n_{23},n_{45}] \left(1+  \frac{a_1 \tilde s_{23}}{F^2}+  \frac{a_2 \tilde s_{23}^2 }{F^4} + \cdots\right) \left(1+  \frac{a_1 \tilde s_{45}}{F^2}+  \frac{a_2 \tilde s_{45}^2 }{F^4} + \cdots\right) \ .
\end{equation}
The 5-point amplitude built up from this cubic interaction is,
{\small
\begin{equation}
\begin{aligned}
\mathcal{A}_{5}[12345] &=\frac{\tilde V_{3}[n_1,n_2,n_{12}] \dbtilde V_{3}[n_{12},n_{34},n_5] \tilde V_{3}[n_3,n_4,n_{34}]}{\tilde{s}_{12} \tilde{s}_{34}}+\frac{\tilde V_{3}[n_5,n_1,n_{15}] \dbtilde V_{3}[n_{15},n_{23},n_4] \tilde V_{3}[n_2,n_3,n_{23}]}{\tilde{s}_{15} \tilde{s}_{23}} \\
&+\frac{\tilde V_{3}[n_1,n_2,n_{12}]\dbtilde V_{3}[n_{45},n_{12},n_3] \tilde V_{3}[n_4,n_5,n_{45}]}{\tilde{s}_{12} \tilde{s}_{45}}+\frac{\tilde V_{3}[n_5,n_1,n_{15}]\dbtilde V_{3}[n_{34},n_{15},n_2]\tilde  V_{3}[n_3,n_4,n_{34} ]}{\tilde{s}_{15} \tilde{s}_{34}} \\
&+\frac{\tilde V_{3}[n_2,n_3,n_{23}]\dbtilde V_{3}[n_{23},n_{45},n_1] \tilde V_{3}[n_4,n_5,n_{45}]}{\tilde{s}_{23} \tilde{s}_{45}}  \ .
\end{aligned}
\end{equation}
}
We start with the 5-point KK relation,
\begin{equation}
\label{eq,5pointKK}
    \mathcal{A}_{5}[12345]+\mathcal{A}_{5}[12354]+\mathcal{A}_{5}[12435]+\mathcal{A}_{5}[14235]=0~.
\end{equation}
At $\mathcal{O}\left(p^{-4}\right)$, one can see that as long as $V_3[i,j,ij]$ is cyclic and anti-symmetric, the 5-point KK relation is automatically satisfied. We now turn to the 5-point BCJ relations which correspond to 4 independent constraints on the color-ordered amplitudes. Since in the multi-scalar theory under consideration, all external particles are scalars in the adjoint of $U(N)$, one can exchange external states by simply relabeling their masses. Thus, the BCJ relations only lead to 2 new constraints on the 5-point amplitudes. The constraint arising from the fundamental BCJ relation reads
\begin{equation}
   \tilde s_{12} \mathcal{A}_5[21345]- \tilde  s_{23} \mathcal{A}_5[13245]-\left(\tilde  s_{23}+\tilde  s_{24}\right) \mathcal{A}_5[13425]=0 \ ,
\end{equation}
and is automatically satisfied after using the 4-point BCJ relation, implementing momentum conservation, and using the mass spectral condition. The other constraint arising from the BCJ relations is
\begin{equation}
    \mathcal{A}_{5}[12453] =\frac{-\mathcal{A}_{5} [12345] \tilde s_{34}\tilde s_{15}-\mathcal{A}_{5} [12354] \tilde s_{14}\left(\tilde s_{13}+\tilde s_{35}\right)}{\tilde s_{24} \tilde s_{13}}~,
\end{equation}
which gives us a new constraint on the cubic interaction: 
\begin{align}
  &-V_3[n_1,n_3,n_{13}] V_3[n_2,n_4,n_{24}] V_3[n_{13},n_{24},n_5] + V_3[n_1,n_4,n_{14}] V_3[n_2,n_5,n_{25}] V_3[n_{14},n_{25},n_3]\nonumber\\
  &+V_3[n_1,n_5,n_{15}] V_3[n_3,n_4,n_{34}] V_3[n_{15},n_{34},n_2]-V_3[n_1,n_2,n_{12}] V_3[n_3,n_5,n_{35}] V_3[n_{12},n_{35},n_4]\nonumber\\
  &- V_3[n_2,n_3,n_{23}] V_3[n_4,n_5,n_{45}] V_3[n_{23},n_{45},n_1]  =0  \ . \label{eq,5ptbcjconstraint}
\end{align} 
This is an extra constraint on the cubic amplitude in addition to the 4-point BCJ relations. We have seen from the analysis of the 4-point amplitude that we require an infinite tower of massive states. Given this, the 5-point BCJ relation, including the corresponding 4-point BCJ relations from all possible sub-graphs and all the spectral conditions arising from these, can always be satisfied, since we can always include a new massive state with new interactions. It is worth pointing out that one can also build the 3-point amplitude using the masses in the theory, i.e. $V_3[1,2,12]^{(1)} = A_{1,2}^{(1)} m_{12}+A_{2,12}^{(1)} m_{1}+A_{12,1}^{(1)} m_{2}$. This simply gives us more free parameters, so that the BCJ relations and spectral conditions can be satisfied for the infinite massive tower as before.

We now check higher-derivative amplitudes. At $\mathcal{O}(p^{-2})$, there are two types of diagrams, one with three cubic vertices, and one with a quartic vertex and a cubic vertex, where the quartic vertex comes from the contact term that is designed to cancel the 4-point amplitude at $\mathcal{O}(p^{-2})$ as in Eq.\eqref{eq,4point1}. The diagram with three cubic vertices is given by
\begin{align}
\mathcal{A}^{(1), 3c}_{5}[12345] &= -  \frac{2 a_1 }{F^2 } V_{3}[n_1,n_2,n_{12}] V_{3}[n_{12},n_{34},n_5] V_{3}[n_3,n_4,n_{34}] \left(\frac{1}{\tilde s_{12} }+\frac{1}{\tilde s_{34} }\right) \nonumber \\
&+ (\text{cyclic permutations}) \ .
\end{align}
Meanwhile, the quartic-cubic diagram gives us
\begin{align}
    \mathcal{A}^{(1), qc}_{5}[12345]& =   \frac{2 a_1   V_3[n_4,n_5,n_{45}]  }{F^2 \tilde s_{45}} \left(V_3[n_1,n_2,n_{12}] V_3[n_3,n_{45},n_{12}] + V_3[n_{45},n_1,n_{23}]V_3[n_2,n_3,n_{23}] \right) \nonumber \\
    &+ (\text{cyclic permutations}) \ ,
\end{align}
which cancels the contributions from the graph with only cubic vertices. This not surprising, since the 4-point contact vertices are constant, so there is no distinction between on-shell and off-shell, and they were constructed so that they would cancel the contributions from these 3-point interactions.

At next-to-next-to leading order (NNLO), the 5-point amplitude reads
\begin{eqnarray}
  \mathcal{A}_{5}^{(2), 3 c}[12345]&=& - \frac{4  a_{1}^2 }{F^{4}} V_{3}\left[n_{1}, n_{2}, n_{12}\right] V_{3}\left[n_{12}, n_{34}, n_{5}\right] V_{3}\left[n_{3}, n_{4}, n_{34}\right]\nonumber\\
  && + (\text{cyclic permutations}) \ .
\end{eqnarray}
This is a constant amplitude that automatically satisfies the 5-point KK relation. However, the 5-point BCJ relation of this amplitude cannot be satisfied. This leads to the need for a new 5-point contact term to cancel it in the same way that we needed a 4-point contact term to cancel the 4-point NLO amplitude in Eq.\eqref{eq,4point1}. At the 6-point level, this will give a null result. Thus, we do not obtain any new contributions, just as in the previous case.

Finally, we want to show that the quartic interaction obtained in Sec.\eqref{sec,quartic} does not mix with the constant cubic interaction. We consider the 5-point amplitude that contains a cubic vertex and the lowest order quartic vertex,
\begin{align}
\label{eq,5pointcubicquartic}
    V_5[12345] &= \frac{V_3[n_1,n_2,n_{12}] \tilde s_{35}}{\tilde s_{12}} +\frac{V_3[n_5,n_1,n_{15}] \tilde s_{24}}{\tilde s_{15}} +\frac{V_3[n_4,n_5,n_{45}] \tilde s_{13}}{\tilde s_{45}} \nonumber \\
    &+\frac{V_3[n_3,n_4,n_{34}] \tilde s_{25}}{\tilde s_{34}} +\frac{V_3[n_2,n_3,n_{23}] \tilde s_{14}}{\tilde s_{23}}  \ .
\end{align}
Substituting this into the 5-point KK relation Eq.\eqref{eq,5pointKK}, we have
\begin{equation}
\label{eq,constraint}
    V_3[n_1,n_2,n_{12}]-V_3[n_1,n_5,n_{15}]+V_3[n_2,n_3,n_{23}]+V_3[n_3,n_5,n_{35}]=0 \ ,
\end{equation}
which can in principle be treated as an extra constraint for the cubic interaction coefficients. Alternatively, one can introduce a contact term 
{\small
\begin{equation}
\label{eq,5pointcontact}
    V_{5c}[12345] = - \frac{1}{3}\left( V_3[n_1,n_2,n_{12}] +V_3[n_5,n_1,n_{15}]+\!V_3[n_4,n_5,n_{45}]+V_3[n_3,n_4,n_{34}]+V_3[n_2,n_3,n_{23}] \right) ,
\end{equation}}so that the KK relation is automatically satisfied. In this case, the constraint in Eq.\eqref{eq,constraint} does not need to hold. One can verify that both options do not satisfy the 5-point BCJ relation. This means that the two interactions are disconnected, and should not exist in the same theory. 
 
\section{Conclusions} 
Beginning with the effective Lagrangians of the 5D massless NLSM and SGal models, we have performed a Kaluza-Klein reduction, compactifying on a circle, to obtain massive 4D scalar theories. By computing the amplitudes of these massive theories, we have found that their amplitudes are constructed in terms of tilded Mandelsatam variables defined in Eq.\eqref{eq,newMandel} that satisfy the same relations as the original Mandelstam variables $s_{ij} = (p_i + p_j)^2$. We have explicitly computed the 4-point and 6-point amplitudes of the NLSM to show that the 4D massive amplitudes are the same as those of the massless theory, with the substitution $s \rightarrow \tilde s$. For the SGal model, since an extra sextic interaction shows up in 5D, the 6-point amplitude in the massive 4D theory has an extra contribution from this contact term. We have shown that this contribution vanishes and that the amplitudes of the 4D massive SGal theory are obtained through simple substitutions from those of the massless theory. These insights, together with dimensional reduction arguments, show that the amplitudes in massive theories obtained from Kaluza-Klein reductions should only depend on the tilded Mandelstam variables. Moreover, in the massive kernel $S[\alpha |\beta]$, one can write the mass pole $s_{ij} - m_{ij}^2 $ as $\tilde s_{ij}$ according to our new definition, which shows that the KLT kernel would have the same rank as that of the massless one, since, again, the algebraic relations among the Mandelstam variables are the same as those for the tilded ones. Just as in the massless theory, null vectors arise and yield BCJ-like relations. We thus conclude that the 4D massive theory obtained from Kaluza-Klein reduction automatically satisfies the mass spectral condition and, therefore, is manifestly local. 

If the mass spectral condition is satisfied, we can then use the tilded Mandelstam variables to build the amplitude. We have constructed new amplitudes by considering a different interaction strength for different massive states and have found that the most general 4-point amplitude from a contact quartic term is the same as the one derived from the Kaluza-Klein reduction. The massive amplitude is then the same as the massless NLSM one after substituting tilded Mandelstam variables for regular Mandelstam variables at all higher derivative orders. 

Interestingly though, these simple relationships change when considering the cubic interaction. In the massless NLSM, there is no cubic interaction. Therefore, the massive theory after Kaluza-Klein reduction will also not contain cubic interactions. However, when exploring the most general amplitude with the mass spectral condition as the only constraint, we are allowed to write down a non-zero cubic interaction. We have proceeded as in the quartic interaction case by writing a cubic interaction with different strength for different massive states. Starting with a cyclic, anti-symmetric 3-point amplitude, the 4-point KK relation is automatically satisfied, and the 4-point BCJ relation imposes an extra constraint on the cubic amplitudes. These cubic interactions lead to a 5-point amplitude whose KK and fundamental 5-point BCJ relations are automatically satisfied, while the other 5-point BCJ relation yields an extra condition. We have verified that such amplitudes can always be written down if we consider an infinite tower of massive states.

Several open questions regarding the massive double copy remain. Within the context of a massive tower of states satisfying the spectral condition, it is interesting to ask whether a dimensional compactification over a topology other than the circle would also give us a lower dimensional theory that automatically satisfies the mass spectral condition. On the other hand, it remains an intriguing question whether the only way to construct a well-defined massive double copy in $D\geq 4$ is to have a tower of massive states with a special spectrum. However, the current proposal relies on using the same structures as in the massless case and such an alternative massive double copy formulation, if it exists, remains elusive in $D\geq 4$.

\section*{Acknowledgements}
MCG thanks Shruti Paranjape for helpful discussions. The work of QL and MT is supported in part by US Department of Energy (HEP) Award DE-SC0013528, and by NASA ATP grant 80NSSC18K0694. The research of MCG is supported by STFC grants ST/P000762/1, ST/T000791/1, and the European Union's Horizon 2020 Research Council grant 724659 MassiveCosmo ERC--2016--COG.

\end{CJK*}
\bibliography{ref}
\end{document}